\documentstyle[prl,twocolumn,aps,epsf]{revtex}
\begin{document}
\draft

\twocolumn[\hsize\textwidth\columnwidth\hsize\csname
@twocolumnfalse\endcsname
\title
 {\bf  Upper  critical  field  in  {Ba$_{1-x}$K$_x$BiO$_3$}:
magnetotransport versus magnetotunneling}

\author{P.  Samuely,$^1$ P.  Szab\'o,$^{1,2}$ T.  Klein,$^3$
  A.G.M.Jansen,$^2$ J.  Marcus,$^3$ C.Escribe-Filippini,$^3$
  and P. Wyder$^2$}

\address{
$^1$Institute of Experimental  Physics, Slovak Academy of
Sciences, SK-04353 Ko\v{s}ice, Slovakia.\\
$^2$Grenoble    High     Magnetic    Field    Laboratory,
Max-Planck-Institut   f\"{u}r   Festk\"{o}rperforschung   and
Centre
National de  la Recherche Scientifique,  B.P. 166, F-38042  Grenoble
Cedex 9, France. \\
$^3$Laboratoire d'Etudes des Propri\'et\'es Electroniques
des Solides, Centre National de la Recherche Scientifique, B.P. 166,
F-38042 Grenoble Cedex 9, France. }

\date{\today}
\maketitle
\begin{abstract}

Elastic  tunneling  is  used  as  a  powerful direct tool to
determine  the  upper  critical  field  $H_{c2}(T)$  in  the
high-$T_c$  oxide Ba$_{1-x}$K$_x$BiO$_3$.
The temperature
dependence of $H_{c2}$ inferred from the tunneling follows the
Werthamer-Helfand-Hohenberg    prediction     for    type-II
superconductors.
A comparison will be made with resistively determined critical field data.
\end{abstract}

]

The   upper   critical    field   $H_{c2}$   of   high-$T_c$
superconductors remains a  contradictory issue. In classical
type-II superconductors this quantity has been unequivocally
determined  from  the  magnetotransport  measurement and its
temperature dependence can, in most cases, be well described
by the Werthamer-Helfand-Hohenberg  (WHH) theory \cite{whh}.
However,  for  the  high-$T_c$  superconductors,
$H_{c2}$ extracted from magnetotransport data
reveals  an  unusual  increase  with decreasing temperatures
without any saturation down to low temperatures. This upward
curvature   is  observed   in  a   pronounced  way   in  the
superconducting       cuprates       Sm$_{2-x}$Ce$_x$CuO$_4$
\cite{sumarlin92},              Tl$_2$Ba$_2$CuO$_{6+\delta}$
\cite{mackenzie93},   Bi$_2$Sr$_2$CuO$_y$  \cite{osofsky93},
and                  YBa$_2$(Cu$_{1-x}$Zn$_x$)O$_{7-\delta}$
\cite{walker95}. However,  the effect is  also found in  the
fully     three     dimensional     and     non     magnetic
Ba$_{1-x}$K$_x$BiO$_3$   \cite{affronte}.    Among   others,
a bipolaron scenario  \cite{alexandrov93}, an unconventional
normal   state  \cite{dias94},   a  strong   electron-phonon
coupling   \cite{marsiglio87},    the presence   of   inhomogeneities
and magnetic impurities \cite{ovchinnikov96}
 have  been put forward
for an explanation of  the anomalous $H_{c2}(T)$ dependence.
  Because  depinned vortices,
either  in  the  liquid  or  solid  state,  cause  a  finite
dissipative resistance  before reaching the  full transition
to the  normal state, the  complexity of the  $H - T$  phase
diagram  in  high-$T_c$'s  \cite{blatter94}  undermines  any
direct determination
of the  upper critical field  from magnetotransport data.
There   are
indications   that,  also   in  the   fully  3D   system  of
Ba$_{1-x}$K$_x$BiO$_3$, fluctuations
can   lead  to   a   melting   of  the   vortex-glass  state
\cite{klein}. This could be a reason complicating a determination
of    $H_{c2}$   from    a   dissipative    measurement   as
magnetoresistance.

Avoiding   the   dissipative    mechanisms   which  could obscure
a determination of $H_{c2}$  from transport measurements, we
show that tunneling measurements can be used as an effective
and  direct  method  for  the  determination  of  the  upper
critical  field.  The  from  tunneling  obtained temperature
dependence   of   $H_{c2}$   in   Ba$_{1-x}$K$_x$BiO$_3$  follows
the WHH model  revealing a saturation at low
temperatures.
At the tunneling $H_{c2}$ values the resistance is
very close to the full resistive transition into the normal state
as measured on the same sample.
This  result was obtained repeatedly  on several
samples.

The single-crystalline  Ba$_{1-x}$K$_x$BiO$_3$ samples grown
by  electrochemical crystallisation  were dark-blue crystals
of  a  cubic   shape  with  a  size  of   about  0.6  mm.  The
superconducting  transition  of   our  crystals  was  single
stepped  and sharp  with $T_c\simeq  23$~K as  determined by
susceptibility measurements. The
low temperature resistivity was about 100 $\mu \Omega
$cm with the metallic temperature dependence. The
tunnel junctions were prepared by  painting a silver spot of
about 0.1 to 0.2~mm diameter  on the surface of the crystal.
The interface between  the silver and Ba$_{1-x}$K$_x$BiO$_3$
counter  electrodes  served  as  a  natural  barrier forming
a planar   normal-insulator-superconductor   (N-I-S)  tunnel
junction. Low  resistance electrical contacts  were prepared
for  the  four-probe  measurements  of  the  current-voltage
($I$-$V$)    and    differential    conductance    ($dI/dV$)
characteristics  of  the   tunnel  junction.  The  tunneling
measurements were  performed in magnetic  fields up to  30~T
perpendicular to the planar  junction enabling the formation
of  the  vortex  state  in  the  junction  area. On the same
samples    the   magnetoresistance    was   measured   using
a four-probe measurement at low frequencies.

Figure   1  shows   the  magnetoresistance   $R(H)$  of  the
Ba$_{1-x}$K$_x$BiO$_3$   single  crystal   up  to   26~T  at
temperatures from 1.5~K to $T_c \simeq 23 $~K. The resistive
transitions   are  shifted   and  broadened   towards  lower
temperatures  as the  magnetic field  is increased, although
the  broadening is  much smaller   than in  the case  of the
cuprates.  A  simple  evaluation  of  the transition field
defined as $H^{\ast}(T)$ where $R/R_n$ equals, for instance,
0.1,  0.5, or  0.9  ($R_n$  is the  normal-state resistance)
leads to  a positive curvature of  $H^{\ast}(T)$ down to the
lowest  temperatures.
If  $H^{\ast}(T)$ is  defined for  $R$ even
closer  to  $R_n$,
the  curvature  of $H^{\ast}(T)$ changes at  the  lowest
temperatures  which makes that the dependence $H^{\ast}(T)$
depends on the chosen criterion.

In  the following  the normalized  tunneling conductances of
our   junctions  are   presented,  where   the  normal-state
conduction for the normalization  is taken above $H_{c2}(T)$
for the temperature $T$  under investigation. Figure 2 shows
a quality certificate  of our junction.  The spectra \begin{figure}[t]
\epsfverbosetrue
\vspace{-35mm}
\epsfxsize=7.5cm
\epsfysize=6cm
    \hspace{-5mm}
\begin{center}
  \epsffile{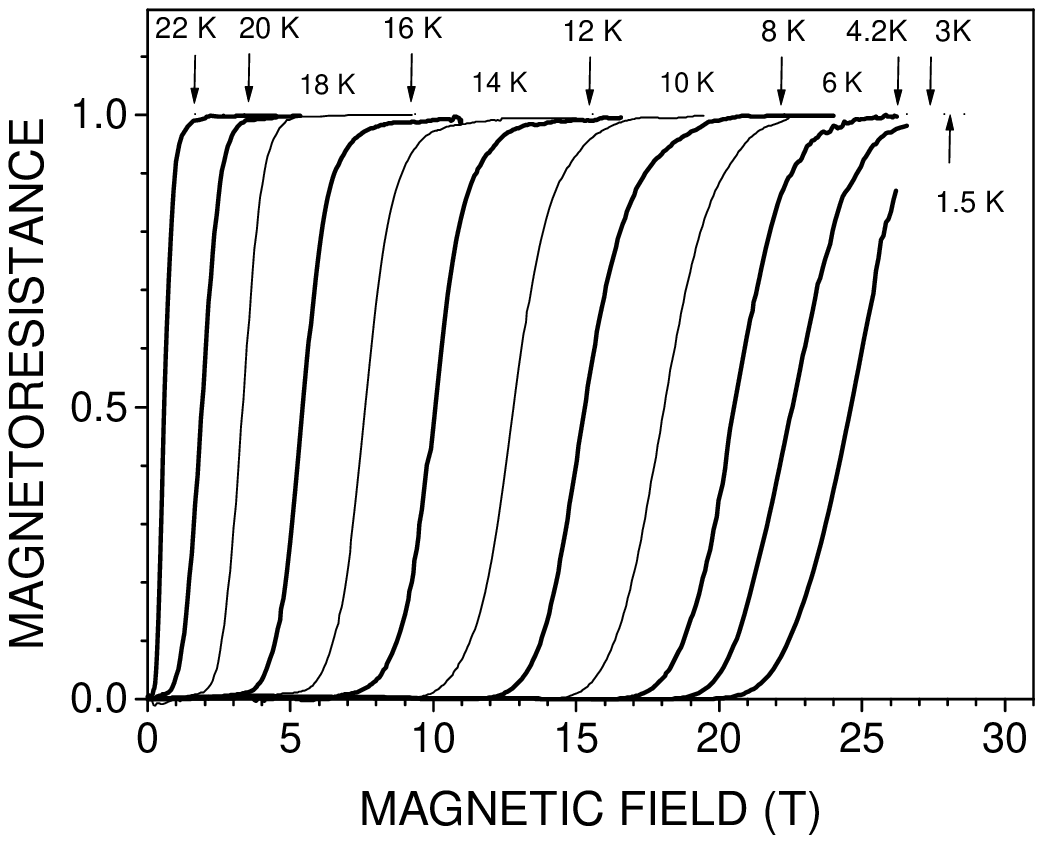}
 \end{center}
\caption{Magnetoresistance   of   the   Ba$_{1-x}$K$_x$BiO$_3$  single
crystal at different temperatures. Arrows - $H^t_{c2}$
obtained from tunneling.}
\end{figure}  can  be
perfectly described by the Dynes formula \cite{dynes} of the
quasi-particle density  of states $\rho  (\epsilon)$ smeared
by the  finite temperature at  which the N-I-S  junction has
been       measured.       $\rho(\epsilon)       =      {\rm
Re}[\epsilon'/(\epsilon'^2  - \Delta^2)^{1/2}]$  contains an
isotropic superconducting energy gap  $\Delta$ and a complex
energy $\epsilon' = \epsilon  - i\Gamma$ which takes account
for some additional smearing $\Gamma$.     The $\Gamma$ smearing
of  the   spectrum  is  case   dependent  with  a   tendency
$\Gamma/\Delta  \rightarrow  0$  in  the  best junctions. As
mentioned  already in  the original  paper of  Dynes et al.,
such  an  "intrinsic"  width  of  the  spectrum  can  be the
consequence of  anisotropy effects, noise,  or concentration
fluctuations.   The   presence   of   microphases   due   to
fluctuations in the potassium and oxygen concentration seems
to  be  a  general  problem  in  a  substitution system like
Ba$_{1-x}$K$_x$BiO$_3$  \cite{affronte}.  The  Dynes formula
fits  our tunnel  spectra at  1.5~K and  zero magnetic field
with  $\Delta  =  3.9  \pm  0.1$~meV  and  $\Gamma = 0.4 \pm
0.1$~meV,  yielding  $2\Delta  /kT_c  =  3.9  \pm  0.1$  and
indicating   that  Ba$_{1-x}$K$_x$BiO$_3$   is  a   BCS-like
superconductor    with    a    medium    coupling   strength
\cite{huang}. In the inset the temperature dependence of the
superconducting  gap is  shown for  the data  from Fig.2 and
also  for  data  taken  at  $B  =  2$~T. The broadening parameter
$\Gamma$  is  found  to  be  independent  of  temperature.

At   high   magnetic   fields   the   normalized   tunneling
conductances  of the  Ba$_{1-x}$K$_x$BiO$_3$-Ag junction are
displayed in Fig.3 for different constant temperatures. With
increasing  magnetic fields  an increasing  smearing of  the
superconducting  features   in  the  tunneling   spectra  is
observed.    At a  certain  field  strength no  structure from
superconductivity  can be  found anymore  and the transition
from a S-I-N to a N-I-N junction is accomplished.
Similar tunneling  data have
been  obtained on  a thin  Ba$_{1-x}$K$_x$BiO$_3$ film  in a
parallel field up to 7 Tesla at 0.45~K
\cite{roesler93}. Unlike a parallel-field configuration,
our experiment on a cubic single crystal involves the \begin{figure}
\epsfverbosetrue
\vspace{35mm}
\epsfxsize=7.5cm
 \epsfysize=6cm
     \hspace{-2mm}
  \begin{center}
\epsffile{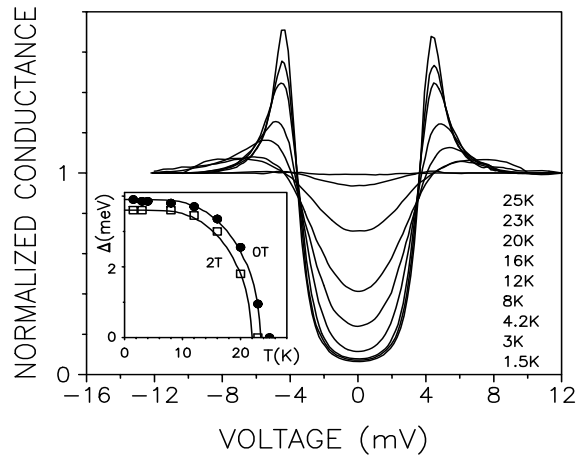}
 \end{center}
\caption{Differential
conductance of the  Ba$_{1-x}$K$_x$BiO$_3$-Ag tunnel junction
measured
at different  temperatures. The inset shows the temperature
dependence of
the  superconducting  energy  gap  obtained  from  the
tunneling conductances at zero magnetic  field and 2~T together with
the
BCS prediction (full lines).}
\end{figure}occurrence
 of a mixed state in a strong magnetic field.

In the  mixed state the tunneling  conductance will probe an
average of  the local densities  of states. For  an isolated
vortex, the  superconducting order parameter is  zero at the
center,  increases  linearly  up  to  a coherence-length distance
$\xi$
away from  the center where  it saturates to  the zero-field
value.  The  local  quasi-particle  density  of states (DOS)
equals  the  normal-state  DOS  at  the  vortex core, but is
broadened near the vortex due to the pair-breaking effect of
the   local   magnetic   field    (as   described   by   the
Abrikosov-Gor'kov  theory  developed  further  by  Maki,  de
Gennes and others \cite{tinkham}).  In the limit of moderate
fields  ($H  <<  H_{c2}$),  Caroli,  de  Gennes and Matricon
\cite{caroli} have  shown that the main  contribution to the
density  of states  at the  Fermi energy  comes from the low
lying  states  localized  in  a  vortex  core. Each isolated
vortex gives a contribution equivalent to a normal region of
radius $\xi$ yielding for the density of states at the Fermi
level $\rho(0) \propto \rho_n(0)  \xi ^2$, where $\rho_n(0)$
is the normal state DOS at  the Fermi level. Thus, the total
averaged   density  of   states  at   the  Fermi   level  is
proportional to $\rho_n(0) \xi  ^2$ per area $(H_{c2}/H) \xi
^2$  occupied   by  one  vortex  giving   $\rho  (0)  \simeq
\rho_n(0) H/H_{c2}$. However, of more relevance for critical
field data, also close to $H_{c2}$ a linear field dependence
of  $\rho(0)$  has  been  found  by  solving  the linearized
Ginzburg-Landau  equation in  the mixed  state \cite{guyon}.
A very sensitive method to determine
the upper  critical field from  tunneling experiments is  to
display  the normalized  zero-bias tunneling  conductance as
a function   of  the   field  strength   \cite{levine}.  The
observation of a  sharp transition is then taken  as a proof
of a good homogeneity of the sample.

In Fig. 4 we present  the zero-bias tunneling conduc-
  \begin{figure}
\epsfverbosetrue
\vspace{10mm}
\epsfxsize=7.5cm
\epsfysize=8.5cm
    \hspace{-3mm}
  \begin{center}
\epsffile{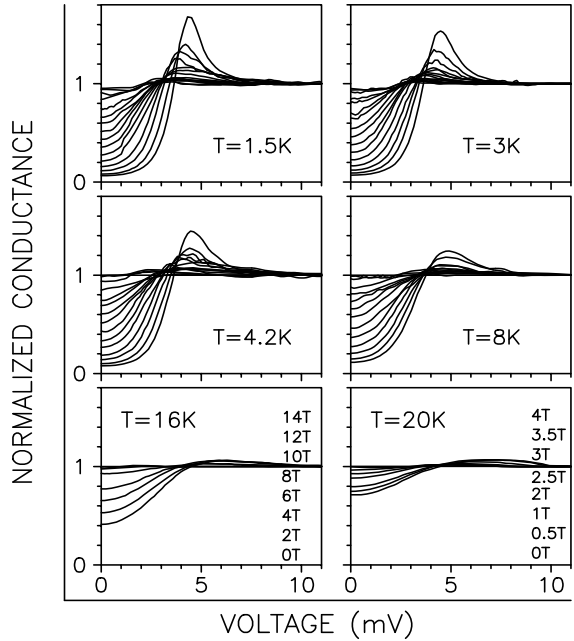}
 \end{center}
\caption{Normalized
tunneling  conductances in  magnetic fields  from zero  up to 30 T
in steps of 2~T (if not mentioned  else) at the indicated
temperatures.}
\end{figure}
tance as
a function  of  the  applied  magnetic  field  for different
temperatures.   A   linear   dependence   of  $dI/dV(0)$  as
a function of applied field
 can  be found in a limited field
range. At  the highest fields  a "tailing" of  the zero-bias
conductance towards the normal-state  value is observed, and
a finite value of the zero-bias conductance is found already
at zero  magnetic field. The latter  effect is obviously due
to the  $\Gamma$ broadening. Also the  tailing effect at the
highest  fields could  be related  to the  same cause as the
$\Gamma$  broadening, i.e.  a certain  inhomogeneity in  the
sample. The  observed tailing effect  resembles the behavior
in the resistive transition close to the transition into the
normal state (see  Fig.1).
A linear extrapolation  of the zero-bias  conductance to the
field  where  the  normalized  conductance  equals unity, as
shown by the full lines in Fig.4, has been used to determine
the  upper  critical  field. Figure  5  shows the
obtained temperature dependence of  the upper critical field
$H^t_{c2}(T)$. Also  the
points  $H^{\ast}(T)$ obtained from  90  \% of the magnetoresistance
transition ($R/R_n = 0.9$) are  indicated.
The  slope of
$H^t_{c2}(T)$  near  $T_c$  is   different  from  that of the
transition field  $H^{\ast}(T)$ determined from  resistance data.
We note that the tunneling critical fields $H^t_{c2}(T)$
are at fields where the bulk resistivity
is very close to the onset of superconductivity
(as indicated by the arrows for $H^t_{c2}$ in Fig.1).

As a very significant result,
$H^t_{c2}(T)$ shows  a clear saturation at  the lowest temperatures
as expected for the WHH theory.
From   all   dissipative   measurements  on
different  samples  of  Ba$_{1-x}$K$_x$BiO$_3$  so  far done
\cite{affronte,klein,roesler93},   \begin{figure}
\epsfverbosetrue
\vspace{10mm}
\epsfxsize=7.5cm
\epsfysize=6cm
    \hspace{-5mm}
  \begin{center}
\epsffile{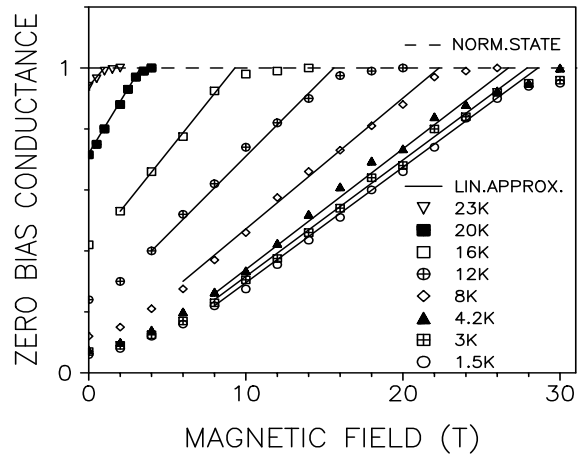}
 \end{center}
\caption{Zero-bias tunneling conductances  as a function  of magnetic
field for all measured temperatures with the linear extrapolation
to obtain $H_{c2}^t$.}
\end{figure}
only a linear increase of  $H^{\ast}(T)$  for decreasing temperatures
has been obtained.
Also  recent
susceptibility  measurements
reveal  this   effect  in  the  temperature   dependence  of  the
irreversibility field  \cite{goll,gantmakher} down to  the lowest
temperatures (0.4 K in \cite{goll}).

Besides the above mentioned tailing effect
in the zero-bias tunneling-conductance and the bulk resistive
transition near the superconducting transition, the same tailing can also be observed near $T_c$
in $\Delta (T)$ (see Fig.2) and  in  $H^t_{c2}(T)$
(see below in  Fig. 5). We suppose that, despite
the quality of our sample, stochiometric  inhomogeneity
could play  a role in this phenomenon.
However, as we will discuss below, a more intrinsic cause
related to superconducting fluctuations
could also explain this broadening in the superconducting transition.

Klein
et  al.  \cite{klein}  suggested  that  in Ba$_{1-x}$K$_x$BiO$_3$
a vortex-glass melting
transition driven by fluctuations
can obscure the magnetotransport
determination of $H_{c2}$.
Their  measurements  of the electric
field versus current density $E-J$  show that a second order
phase  transition from  a vortex  glass to  a vortex  liquid
state does exist in this  system. The presence of the liquid
phase can induce strong fluctuations  below $H_{c2}$ related to
the motion  of the flux  lines, but these  fluctuations
can be quite small above $H_{c2}$.  This can be the reason why
$H_{c2}$ (see arrows in Fig.1)
is  quite  close  to  the  onset of the resistive
transitions . In this approach the foot
of the resistive transition is  determined by the melting of
the  vortex  lattice  giving  a  positive  curvature  in the
temperature dependence of the line $H_g(T)$ for the
liquid-solid transition.  The foots of the  curves in Fig.1
could  be indeed well  fitted  as  $R  \sim  [H/H_g(T)-1]^{\beta}$
corresponding  to   the  vortex  glass   melting  theory  as
introduced by Fisher et al. \cite{fisher}, where $H_g(T)$ is
the magnetic field of  the melting transition. The resulting
fitting  parameter $\beta  = 4.1  \pm 0.5$     \begin{figure}
\epsfverbosetrue
\vspace{10mm}
\epsfxsize=8.2cm
\epsfysize=6cm
    \hspace{-5mm}
  \begin{center}
\epsffile{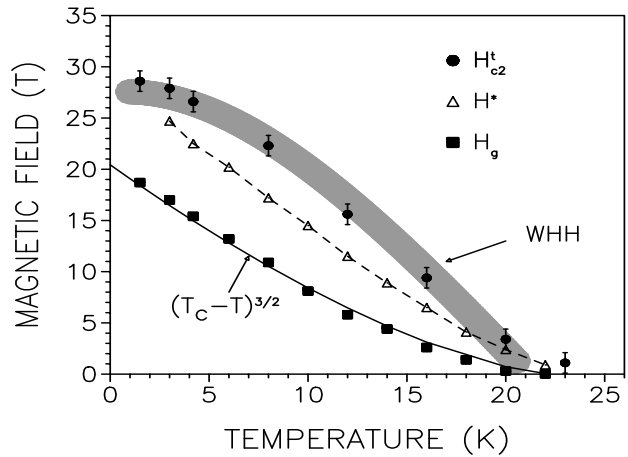}
 \end{center}
\vspace{2mm}
\caption{$H$-$T$  phase
diagram  of Ba$_{1-x}$K$_x$BiO$_3$.  Upper closed  circles -
$H^t_{c2}$
from tunneling,  open triangles - $H^{\ast}$ from
magnetoresistance transition at $R/R_n=0.9$, and closed squares - the
melting line $H_g$.}
\end{figure}
 is in  a perfect
agreement   with   the   value   obtained   on  a  different
Ba$_{1-x}$K$_x$BiO$_3$   crystal   with   $T_c   \simeq$  31
K \cite{klein}. The melting line $H_g(T)$ reveals a positive
curvature and can be  described by the power-law temperature
dependence  $H_g =  H_g(0)(1-T/T_c)^{3/2}$ as  shown in Fig.
5. This is also in a good agreement with  the recent measurement
of   the   irreversibility   field   on   a  similar  sample
\cite{goll} indicating that  the melting and irreversibility
lines coincide in Ba$_{1-x}$K$_x$BiO$_3$.

In the $H - T$ phase-diagram of Fig.5 the initial slope of the
upper  critical  field  $(-dH_{c2}/dT)_{T_c}$  is  about 1.7
$\div $ 1.8 T/K.
To emphasize  more the fact  of the saturation  of the upper
critical field at the lowest temperatures, we note the
closeness of  the zero-bias conductance data  for 1.5, 3 and
4.2~K   in  comparison   with  the   data  taken   at  other
temperatures  in Fig.4.  In Fig.5  we also  present the  WHH
upper critical field line  with an uncertainty comparable to
the  experimental error  bars. Taking  into account  that in
a system with  important fluctuations the  $H_{c2}$ boundary
should not be very sharp a satisfactory agreement is found.

We  have  presented  here a
direct non-dissipative  determination of the  upper critical
field in Ba$_{1-x}$K$_x$BiO$_3$ using  the tunneling effect.
$H_{c2}(T)$ can be
satisfactorily described  by the Werthamer-Helfand-Hohenberg
theory.  In the Cu-oxides, the existence  of a resistive state
within  a large  part of  the $H-T$  diagram
complicates an unambiguous determination
of the critical field from transport data.
Therefore, it would  be very interesting (and
decisive for certain proposed superconducting mechanisms) to
study   the   upper   critical    field  in the cuprates  with   the
non-dissipative tunneling method.

\acknowledgements

We acknowledge  fruitful discussions with  S.I. Vedeneev and
a support of  the EU grant No.CIPA-CT93-0183  and the Slovak
VEGA contract No. 2/1357/94.

\end{document}